\documentclass[11pt]{article}
\usepackage{epsfig}
\usepackage{amsmath,amsfonts}

\begin{document}

\title{\textbf{Interpolating among the Landau, Coulomb and maximal Abelian gauges} }
\author{\textbf{M.A.L. Capri$^{a}$\thanks{%
marcio@dft.if.uerj.br}} \ , \textbf{R.F. Sobreiro}$^{a}$\thanks{%
sobreiro@uerj.br} \ , \textbf{S.P. Sorella}$^{a}$\thanks{%
sorella@uerj.br}{\ }\footnote{Work supported by FAPERJ, Funda{\c
c}{\~a}o de Amparo {\`a} Pesquisa do Estado do Rio de Janeiro,
under the program {\it Cientista do Nosso Estado},
E-26/151.947/2004.} \\\\
\textit{$^{a}$\small{Departamento de F\'{\i }sica Te\'{o}rica}}\\
\textit{\small{Instituto de F\'{\i }sica, UERJ, Universidade do Estado do Rio de Janeiro}} \\
\textit{\small{Rua S{\~a}o Francisco Xavier 524, 20550-013 Maracan{\~a}}} \\
\textit{\small{Rio de Janeiro, Brasil}}}

\maketitle

\begin{abstract}

\noindent A generalized gauge fixing which interpolates among the
Landau, Coulomb and maximal Abelian gauges is constructed.

\end{abstract}

\newpage

\section{Introduction}

The aim of this letter is that of pointing out the possibility of
introducing a generalized gauge fixing which interpolates among
the Landau, Coulomb and maximal Abelian gauges. These gauges have
been employed intensively through theoretical analysis and lattice
numerical simulations in order to investigate several aspects of
Yang-Mills theories in the infrared region. A partial list of them
is given by\footnote{See the reviews
\cite{Greensite:2003bk,Alkofer:2000wg} and references therein.}:
\begin{itemize}
\item  study of the Gribov copies and their influence on the gluon
and ghost propagators,

\item  analysis of Yang-Mills theories through the Schwinger-Dyson
equations,

\item  study of renormalization group invariant effective
couplings and their behavior in the low energy region,

\item  dual superconductivity mechanism for color confinement,

\item  dimension two gauge condensates and their relevance for the
infrared dynamics of gauge theories.
\end{itemize}
\noindent To some extent, this interpolating gauge can be seen as
a generalization of previous results in which suitable gauge
fixings interpolating between the Landau and the Coulomb gauges
\cite{Baulieu:1998kx} as well as between the Landau and the
maximal Abelian gauges \cite{Dudal:2004rx} have been obtained and
proven to be renormalizable to all orders of perturbation
theory.\\\\As we shall also introduce a generalized interpolating
dimension two mass operator, it is worth here spending a few words
on the issue of the gauge invariance of the dimension two
condensates, a topic which is still under debate. Although the
dimension two operator $\int d^{4}xA^{2}$ is not left invariant
by the gauge transformations, it enjoys the property of being $BRST\;$%
invariant on-shell in the Landau, Curci-Ferrari and maximal
Abelian gauges\footnote{In the case of the Curci-Ferrari and
maximal Abelian gauge a slightly more
general operator has to be considered, namely $\left( \frac{A^{2}}{2}+\alpha \overline{%
c}c\right) $.}. This property has made possible to prove that the
operator $ A^{2}$ is multiplicatively renormalizable to all orders
in all these gauges, refs.\cite{Dudal:2002pq,Dudal:2003pe}, a
feature which has been extended to the case of the more general
class of the linear covariant gauges \cite{Dudal:2003np}. This has
enabled us to construct a renormalizable effective potential for
the composite operator $A^{2}$ and to investigate its condensation
in all the aforementioned gauges, see
refs.\cite{Verschelde:2001ia,Dudal:2003vv,Browne:2003uv,Dudal:2003gu,Dudal:2003by,Dudal:2004rx},
providing evidence of a nonvanishing condensate, i.e.
$\left\langle A^{2}\right\rangle \neq 0$, resulting in an
effective gluon mass, $m_{eff}\propto \left\langle
A^{2}\right\rangle $. Moreover, the output of our calculations
shows that the condensate itself is not gauge invariant, {\it
i.e.} the effective gluon mass $m_{eff}$ depends on the gauge
parameter, see for example ref.\cite{Dudal:2003by} where the case
of the linear covariant gauge has been considered. Notice that,
due to color confinement, gluons cannot be observed as free
particles, so that the effective gluon mass cannot be associated
to a quantity which can be directly observable. Nevertheless, a
well defined gauge invariant quantity can be introduced and
computed order by order, and this in the presence of a
nonvanishing condensate $\left\langle A^{2}\right\rangle \neq 0$.
This quantity is the vacuum energy of the theory, $E_{vac}$, its
physical meaning being, of course, apparent. The construction and
the computation of the vacuum energy $E_{vac}$ in the presence of
a nonvanishing condensate $\left\langle A^{2}\right\rangle $, as
well as its independence from the gauge parameters can be found in
\cite{Verschelde:2001ia,Dudal:2003vv,Browne:2003uv,Dudal:2003gu,Dudal:2003by,Dudal:2004rx}.
\\\\In this context, the possibility of having at our disposal a
generalized gauge which interpolates among the Landau, Coulomb and
maximal Abelian gauges might be very useful. It could allow us to
give a proof of the fact that the vacuum energy of the theory has
to be same for all three gauges. \\\\The present work is organized
as follows. In sect.2 we briefly remind some basic properties of
the maximal Abelian gauge. Sect.3 is devoted to the introduction
of the aforementioned interpolating gauge. A suitable dimension
two operator which turns out to be $BRST$\ invariant on-shell and
which interpolates between the dimension two gluon mass operators
already studied in the Landau, Coulomb and maximal Abelian gauges
is introduced in Sect.4. A few remarks on possible further
analytic studies and lattice numerical investigations will be
outlined in the conclusion.

\section{The maximal Abelian gauge and its nonrenormalization theorem}

In order to introduce the maximal Abelian gauge let us briefly fix
the notation. According to \cite{Dudal:2004rx}, we decompose the
gauge field $A_{\mu }^{A}$, $A=1....N^{2}-1$, into off-diagonal
and diagonal components, namely
\begin{equation}
A_{\mu }^{A}T^{A}=A_{\mu }^{a}T^{a}+A_{\mu }^{i}T^{i}\;,  \label{mag1}
\end{equation}
where $T^{A}$ are the generators of the gauge group $SU(N)$,
$\left[ T^{A},T^{B}\right] =if^{ABC}T^{C}$. The indices $i,j,....$
label the $N-1$ diagonal generators of the Cartan subalgebra. The
remaining $N(N-1)$ off-diagonal generators will be labelled by the
indices $a,b,..$. \\\\For the nilpotent $BRST\;$transformations of
the fields, we have
\begin{eqnarray}
sA_{\mu }^{a}\!\!\!\! &=&\!\!\!\!-(D_{\mu
}^{ab}c^{b}+gf^{abc}A_{\mu }^{b}c^{c}+gf^{abi}A_{\mu
}^{b}c^{i})\;,  \nonumber \\
sA_{\mu }^{i}\!\!\!\! &=&\!\!\!\!-(\partial _{\mu
}c^{i}+gf^{abi}A_{\mu
}^{a}c^{b})\;,  \nonumber \\
sc^{a}\!\!\!\! &=&\!\!\!\!gf^{abi}c^{b}c^{i}+\frac{g}{2}f^{abc}c^{b}c^{c}\;,
\nonumber \\
sc^{i}\!\!\!\! &=&\!\!\!\!\frac{g}{2}f^{abi}c^{a}c^{b}\;,  \nonumber \\
s\bar{c}^{a}\!\!\!\! &=&\!\!\!\!b^{a},\qquad sb^{a}=0\;,  \nonumber \\
s\bar{c}^{i}\!\!\!\! &=&\!\!\!\!b^{i},\qquad sb^{i}=0.  \label{mag2}
\end{eqnarray}
where $\left( c^{i},\bar{c}^{i}\right) $, $\left( c^{a},\bar{c}^{a}\right) $
stand for the diagonal and off-diagonal Faddeev-Popov ghosts, while $\left(
b^{i},b^{a}\right) $ denote the diagonal and off-diagonal Lagrange
multipliers. The covariant derivative $D_{\mu }^{ab}$ in eq.$\left( \ref
{mag2}\right) $ is defined as
\begin{equation}
D_{\mu }^{ab}=\delta ^{ab}\partial _{\mu }-gf^{abi}A_{\mu }^{i}\;.
\label{mag3}
\end{equation}
Also, for the field strength one gets $F_{\mu \nu }^{A}=(F_{\mu
\nu }^{i},F_{\mu \nu }^{a})$, {\it i.e.}
\begin{eqnarray}
F_{\mu \nu }^{a}\!\!\!\! &=&\!\!\!\!D_{\mu }^{ab}A_{\nu }^{b}-D_{\nu
}^{ab}A_{\mu }^{b}+gf^{abc}A_{\mu }^{b}A_{\nu }^{c}\;,  \label{mag4} \\
F_{\mu \nu }^{i}\!\!\!\! &=&\!\!\!\!\partial _{\mu }A_{\nu }^{i}-\partial
_{\nu }A_{\mu }^{i}+gf^{abi}A_{\mu }^{a}A_{\nu }^{b}\;.  \nonumber
\end{eqnarray}
Thus, for the Yang-Mills action, \noindent $S_{YM}$, quantized in
the maximal Abelian gauge, $S_{MAG}$, we have
\begin{equation}
S_{YM}+S_{MAG}\;,  \label{qym}
\end{equation}
with
\begin{equation}
S_{YM}=\frac{1}{4}\int d^{4}xF_{\mu \nu }^{A}F_{\mu \nu }^{A}=\frac{1}{4}%
\int d^{4}x\left( F_{\mu \nu }^{a}F_{\mu \nu }^{a}+F_{\mu \nu }^{i}F_{\mu
\nu }^{i}\right) \;,  \label{ym1}
\end{equation}
and
\begin{eqnarray}
S_{MAG} &=&s\int d^{4}x\left( \bar{c}^{a}\partial _{\mu }A_{\mu }^{a}-g\bar{c%
}^{a}f^{abi}A_{\mu }^{i}A_{\mu }^{b}+\frac{\alpha }{2}\bar{c}^{a}b^{a}-\frac{%
\alpha }{2}gf^{abi}\bar{c}^{a}\bar{c}^{b}c^{i}\right.   \nonumber \\
&&\;\;\;\;\;\left. -\frac{\alpha }{4}gf^{abc}c^{a}\bar{c}^{b}\bar{c}^{c}+%
\bar{c}^{i}\partial _{\mu }A_{\mu }^{i}\right) \text{\ ,}  \label{mgf}
\end{eqnarray}
which yields
\begin{eqnarray}
S_{MAG} &=&\int d^{4}x\left( b^{a}\left( D_{\mu }^{ab}A_{\mu }^{b}+\frac{%
\alpha }{2}b^{a}\right) +\overline{c}^{a}D_{\mu }^{ab}D_{\mu }^{bc}c^{c}+g%
\overline{c}^{a}f^{abi}\left( D_{\mu }^{bc}A_{\mu }^{c}\right) c^{i}+g%
\overline{c}^{a}D_{\mu }^{ab}\left( f^{bcd}A_{\mu }^{c}c^{d}\right) \right.
\nonumber \\
&-&\alpha gf^{abi}b^{a}\overline{c}^{b}c^{i}-\left. g^{2}f^{abi}f^{cdi}%
\overline{c}^{a}c^{d}A_{\mu }^{b}A_{\mu }^{c}-\frac{\alpha }{2}gf^{abc}b^{a}%
\overline{c}^{b}c^{c}-\frac{\alpha }{4}g^{2}f^{abi}f^{cdi}\overline{c}^{a}%
\overline{c}^{b}c^{c}c^{d}\right.   \nonumber \\
&-&\left. \frac{\alpha }{4}g^{2}f^{abc}f^{adi}\overline{c}^{b}\overline{c}%
^{c}c^{d}c^{i}-\frac{\alpha }{8}g^{2}f^{abc}f^{ade}\overline{c}^{b}\overline{%
c}^{c}c^{d}c^{e}+b^{i}\partial _{\mu }A_{\mu }^{i}+\overline{c}^{i}\partial
_{\mu }\left( \partial _{\mu }c^{i}+gf\,^{iab}A_{\mu }^{a}c^{b}\right)
\right) \;.  \nonumber \\
&&  \label{mgf1}
\end{eqnarray}
The gauge parameter $\alpha $ in expression $\left(
\ref{mgf}\right) $ has to be introduced for renormalization
purposes \cite{Dudal:2004rx}. The action $\left(\ref{qym}\right)$
displays mutiplicative renormalizability. In particular, we
underline that, as a consequence of the Ward identities which can
be established in the maximal Abelian gauge, the following
nonrenormalization theorem holds, see for instance eq.(40) of
\cite{Dudal:2004rx}, \textit{i.e.}
\begin{equation}
Z_{g}Z_{A^{i}}^{1/2}=1\;.  \label{mgf3}
\end{equation}
This relationship states that the renormalization of the gauge coupling
constant $g$ is related to the renormalization factor of the diagonal
components, $A_{\mu }^{i}$, of the gauge field. Till now, eq.$\left( \ref
{mgf3}\right) $ has been established to all orders of perturbation theory,
being explicitly checked at three loops in \cite{Gracey:2005vu}.

\section{The generalized interpolating gauge}

Following \cite{Baulieu:1998kx}, let us make use of the notation
\begin{eqnarray}
\widetilde{\partial }_{\mu } &=&(\nabla ,a\partial _{4})\;,  \nonumber \\
\widetilde{A}_{\mu }^{i} &=&\left( \overrightarrow{A}^{i},aA_{4}^{i}\right)
\;,  \nonumber \\
\widetilde{A}_{\mu }^{a} &=&\left( \overrightarrow{A}^{a},aA_{4}^{a}\right)
\;,  \nonumber \\
\widetilde{D}_{\mu }^{ab} &=&\delta ^{ab}\widetilde{\partial }_{\mu
}-gf^{abi}\widetilde{A}_{\mu }^{i}\;,  \label{n666}
\end{eqnarray}
where $a$ is the gauge parameter which interpolates between the
Coulomb and Landau gauges. \\\\The gauge fixing, $S_{CLM}$, which
interpolates among the Landau, Coulomb and maximal Abelian gauges
contains three gauge parameters, $\left( a,k,\alpha \right) $,
being given by
\begin{eqnarray}
S_{CLM} &=&s\int d^{4}x\left( \bar{c}^{a}\widetilde{\partial }_{\mu }A_{\mu
}^{a}-g\bar{c}^{a}f^{abi}\widetilde{A}_{\mu }^{i}A_{\mu }^{b}+\frac{\alpha }{%
2}\bar{c}^{a}b^{a}-\frac{\alpha }{2}gf^{abi}\bar{c}^{a}\bar{c}%
^{b}c^{i}\right.   \nonumber \\
&&\;\;\;\;\;\left. -\frac{\alpha }{4}gf^{abc}c^{a}\bar{c}^{b}\bar{c}^{c}+%
\bar{c}^{i}\widetilde{\partial }_{\mu }A_{\mu }^{i}-kgf^{iab}\widetilde{A}%
_{\mu }^{i}A_{\mu }^{a}\overline{c}^{b}\right) \text{\ .}  \label{n6}
\end{eqnarray}
Acting with the $BRST$\ operator $s$ on the elementary fields, one obtains
\begin{eqnarray}
S_{CLM} &=&\int d^{4}x\left( b^{a}\left( \widetilde{D}_{\mu }^{ab}A_{\mu
}^{b}+\frac{\alpha }{2}b^{a}\right) +\overline{c}^{a}\widetilde{D}_{\mu
}^{ab}D_{\mu }^{bc}c^{c}+g\overline{c}^{a}f^{abi}\left( \widetilde{D}_{\mu
}^{bc}A_{\mu }^{c}\right) c^{i}+g\overline{c}^{a}\widetilde{D}_{\mu
}^{ab}\left( f^{bcd}A_{\mu }^{c}c^{d}\right) \right.   \nonumber \\
&-&\alpha gf^{abi}b^{a}\overline{c}^{b}c^{i}-g^{2}f^{abi}f^{cdi}\overline{c}%
^{a}c^{d}\widetilde{A}_{\mu }^{b}A_{\mu }^{c}-\frac{\alpha }{2}gf^{abc}b^{a}%
\overline{c}^{b}c^{c}-\frac{\alpha }{4}g^{2}f^{abi}f^{cdi}\overline{c}^{a}%
\overline{c}^{b}c^{c}c^{d}  \nonumber \\
&-&\frac{\alpha }{4}g^{2}f^{abc}f^{adi}\overline{c}^{b}\overline{c}%
^{c}c^{d}c^{i}-\frac{\alpha }{8}g^{2}f^{abc}f^{ade}\overline{c}^{b}\overline{%
c}^{c}c^{d}c^{e}  \nonumber \\
&&+b^{i}\widetilde{\partial }_{\mu }A_{\mu }^{i}+\overline{c}^{i}\widetilde{%
\partial }_{\mu }\left( \partial _{\mu }c^{i}+gf\,^{iab}A_{\mu
}^{a}c^{b}\right) \;+kgf^{abi}\widetilde{A}_{\mu }^{a}\left(
\partial _{\mu
}c^{i}\right) \overline{c}^{b}  \nonumber \\
&&+kg^{2}f^{abi}f^{cdi}\overline{c}^{a}c^{d}\widetilde{A}_{\mu }^{b}A_{\mu
}^{c}-kgf^{abi}\widetilde{A}_{\mu }^{i}A_{\mu }^{a}\left( b^{b}-gf^{bcj}%
\overline{c}^{c}c^{j}\right)   \nonumber \\
&&\left. +kgf^{abi}\widetilde{A}_{\mu }^{i}\left( D_{\mu }^{ac}c^{c}\right)
\overline{c}^{b}+kg^{2}f^{abi}f^{acd}\widetilde{A}_{\mu }^{i}A_{\mu
}^{c}c^{d}\overline{c}^{b}\right)  \;. \nonumber \\
&&  \label{n66}
\end{eqnarray}
Let us now show how the various gauges can be recovered from expression $%
\left( \ref{n6}\right) $ when appropriate limits for the
parameters $\left( a,k,\alpha \right) $ are taken. Let us begin
with the Landau gauge.

\subsection{The Landau gauge}

The Landau gauge is recovered by taking
\begin{equation}
a=1\;,\;\;\;k=1\;,\;\;\;\alpha =0\;.  \label{n9}
\end{equation}
In fact, from expression $\left( \ref{n6}\right) $, one obtains
\begin{equation}
S_{L}=s\int d^{4}x\left( \bar{c}^{a}\partial _{\mu }A_{\mu }^{a}+\bar{c}%
^{i}\partial _{\mu }A_{\mu }^{i}\right) =s\int d^{4}x\left( \bar{c}%
^{A}\partial _{\mu }A_{\mu }^{A}\right) \;,  \label{n10}
\end{equation}
which is the Landau gauge.

\subsection{The Coulomb gauge}

The Coulomb gauge is obtained from $\left( \ref{n6}\right) $ by setting
\begin{equation}
a=0\;,\;\;\;k=1\,,\,\,\;\,\alpha =0\;.  \label{n7}
\end{equation}
Expression $\left( \ref{n6}\right) $ is easily seen to reduce to the Coulomb
gauge, namely
\begin{equation}
S_{C}=s\int d^{4}x\left( \bar{c}^{a}\left( \nabla \cdot \overrightarrow{A}%
^{a}\right) +\bar{c}^{i}\left( \nabla \cdot \overrightarrow{A}^{i}\right)
\right) =s\int d^{4}x\;\bar{c}^{A}\left( \nabla \cdot \overrightarrow{A}%
^{A}\right) \;.  \label{n8}
\end{equation}

\subsection{The maximal Abelian gauge}

Finally, the maximal Abelian gauge corresponds to
\begin{equation}
a=1\;,\;\;\;k=0\;,  \label{n11}
\end{equation}
yielding

\begin{eqnarray}
S_{MAG} &=&s\int d^{4}x\left( \bar{c}^{a}\partial _{\mu }A_{\mu }^{a}-g\bar{c%
}^{a}f^{abi}A_{\mu }^{i}A_{\mu }^{b}+\frac{\alpha }{2}\bar{c}^{a}b^{a}-\frac{%
\alpha }{2}gf^{abi}\bar{c}^{a}\bar{c}^{b}c^{i}\right.  \nonumber \\
&&\;\;\;\;\;\left. -\frac{\alpha }{4}gf^{abc}c^{a}\bar{c}^{b}\bar{c}^{c}+%
\bar{c}^{i}\partial _{\mu }A_{\mu }^{i}\right) \text{\ .}  \label{n12}
\end{eqnarray}
which is recognized to be the maximal Abelian gauge, eq.$\left( \ref{mgf}%
\right) $.

\section{An interpolating mass dimension two operator}

It is worth remarking that the interpolating gauge fixing $\left( \ref{n6}%
\right) $ allows us to introduce a generalized mass dimension two operator, $%
O_{CLM}$
\begin{equation}
O_{CLM}=\frac{1}{2}\widetilde{A}_{\mu }^{a}A_{\mu }^{a}+\frac{k}{2}%
\widetilde{A}_{\mu }^{i}A_{\mu }^{i}+\alpha \bar{c}^{a}c^{a}\text{\ ,}
\label{n13}
\end{equation}
which enjoys the property of being $BRST\;$invariant on-shell. More
precisely, one has
\begin{equation}
s\int d^{4}xO_{CLM}=\int d^{4}x\;\left( kc^{i}\frac{\delta
(S_{YM}+S_{CLM})}{\delta b^{i}}+c^{a}\frac{\delta
(S_{YM}+S_{CLM})}{\delta b^{a}}\right) \;.  \label{n14}
\end{equation}
Interestingly, the operator $O_{CLM}$ interpolates among all
dimension two mass operators already introduced in the Coulomb,
Landau and maximal Abelian gauges, namely
\begin{equation}
O_{CLM}\rightarrow O_{Coulomb}=\frac{1}{2}\overrightarrow{A}^{A}\cdot
\overrightarrow{A}^{A}\text{\ ,\ \ \ \ \textrm{for \ }}a=0\;,\;\;\;k=1\,,\,%
\,\;\,\alpha =0\;,  \label{n15}
\end{equation}
\begin{equation}
O_{CLM}\rightarrow O_{Landau}=\frac{1}{2}A_{\mu }^{A}A_{\mu }^{A}\text{\ ,\
\ \ \ \textrm{for \ }}a=1\;,\;\;\;k=1\,,\,\,\;\,\alpha =0\;,  \label{n16}
\end{equation}
\begin{equation}
O_{CLM}\rightarrow O_{MAG}=\frac{1}{2}A_{\mu }^{a}A_{\mu }^{a}+\alpha \bar{c}%
^{a}c^{a}\text{\ ,\ \ \ \ \textrm{for \ }}a=1\;,\;\;\;k=0\,\;.  \label{n17}
\end{equation}
Analytic evidence for the condensation of all these mass
dimensions two operators have been given in
refs.\cite{Greensite:1985vq,Verschelde:2001ia,Dudal:2003vv,Browne:2003uv,Dudal:2004rx}.

\section{Conclusion}

In this work a generalized gauge, eq.$\left( \ref{n6}\right) $,
which interpolates among the Landau, Coulomb and maximal Abelian
gauges has been introduced. It could lead to several interesting
features worth to be analysed.

\begin{itemize}
\item  One first aspect to be faced is the renormalizability of
the interpolating gauge $\left( \ref{n6}\right) $. This point is
being investigated along the lines of \cite{Baulieu:1998kx}, where
an all orders $BRST$ algebraic proof of the multiplicative
renormalizability of the gauge fixing interpolating between the
Landau and Coulomb gauges has been achieved. Also, it would be
interesting to see if the algebraic setup of \cite{Baulieu:1998kx}
could be generalized so as to include the dimension two mass
operator $O_{CLM}$, eq.$\left( \ref{n13}\right) $, as well as the
nonlocal gauge invariant operator $Tr\int d^{4}xF_{\mu \nu }\frac{1}{D^{2}}%
F_{\mu \nu }$ recently discussed in \cite{Capri:2005dy} within the
class of the linear covariant gauges, which include the Landau
gauge as particular case. \\\\As already mentioned in the
introduction, although the dimension two operator $A^{2}$ has been
proven to be multiplicatively renormalizable to all orders in the
Landau, Curci-Ferrari, maximal Abelian and linear covariant
gauges, refs.\cite{Dudal:2002pq,Dudal:2003pe,Dudal:2003np}, a
clear understanding of the aspects related to the gauge invariance
of the dimension two condensate, $\langle A^{2} \rangle$, is still
under analysis. As discussed in \cite{Gubarev:2000eu}, a possible
gauge invariant extension of the operator $A^{2}$ could be
provided by the gauge invariant operator $A_{\mathrm{\min }}^{2}$,
obtained by minimizing $A^{2}$ along the gauge orbit of $A_{\mu
}$. However, the operator $A_{\mathrm{\min }}^{2}$ appears to be
highly nonlocal, a feature which jeopardizes the standard
perturbative renormalization procedure for an arbitrary choice of
the gauge fixing \cite{Esole:2004jd}, see also
\cite{Capri:2005dy}. Nevertheless, as shown in
\cite{Dudal:2003gu,Dudal:2003by,Dudal:2004rx}, the vacuum energy,
$E_{\it vac}$, evaluated in the presence of the dimension two
condensate $\langle A^{2} \rangle$ turns out to be independent
from the gauge parameters. Therefore, the possibility of having at
our disposal a generalized gauge which interpolates among the
Landau, Coulomb and maximal Abelian gauges might allow us to
achieve the result that the vacuum energy evaluated in the
presence of the generalized dimension two operator, eq.$\left(
\ref{n13}\right) $, has to be same for all three gauges.

\item  A second aspect which could be exploited is to investigate
whether the nonrenormalization theorem of the maximal Abelian
gauge, as expressed by eq.$\left( \ref{mgf3}\right) $, would
remain valid beyond perturbation theory. The natural framework to
discuss this issue is through lattice numerical simulations as
done, for example,  in the case of the nonrenormalization theorem
of the ghost-gluon vertex in the Landau gauge
\cite{Cucchieri:2004sq}. It is worth underlining that the
relationship $\left( \ref{mgf3}\right)$ could open the interesting
possibility of studying, through lattice simulations, the infrared
behavior of the running coupling
constant in the maximal Abelian gauge. More precisely, eq.$\left( \ref{mgf3}%
\right)$ suggests that the infrared behavior of the running
coupling constant in the maximal Abelian gauge could be accessed
by looking at the behavior of the diagonal component of the gluon
propagator. Interestingly, a recent study of the gluon propagator
in momentum space has been performed in \cite{Bornyakov:2003ee},
reporting an infrared suppression of the diagonal component.
Moreover, according to \cite{Bornyakov:2003ee},
the diagonal gluon propagator could attain a finite nonvanishing value at $%
k\approx 0$, a feature which could signal the possible existence
of an infrared fixed point for the running coupling constant in
the maximal Abelian gauge.

\item Finally, we point out that the authors \cite{Fischer:2005qe}
have shown that the use of the interpolating Landau-Coulomb gauge
allows to introduce two renormalization group invariant running
couplings. In particular, one of these two couplings turns out to
be independent from the interpolating gauge parameter, displaying
an infrared fixed point whose value coincides with that already
known in the Landau gauge. The existence of such an infrared fixed
point extends thus to the Coulomb gauge too \cite{Fischer:2005qe}.
It would be worth to investigate if similar effective
renormalization group invariant couplings could be introduced also
in the present interpolating gauge. This might provide further
indication on the possible existence of an infrared fixed point
for the running coupling constant in the maximal Abelian gauge.
\end{itemize}

\section*{Acknowledgments.}

S.P. Sorella is grateful to D. Zwanziger for useful remarks and to
A. Cucchieri for interesting preliminary discussions about the
possibility of investigating relation $\left( \ref{mgf3}\right)$
through lattice simulations. We thank the Conselho Nacional de
Desenvolvimento Cient\'{\i}fico e Tecnol\'{o}gico (CNPq-Brazil),
the Faperj, Funda{\c c}{\~a}o de Amparo {\`a} Pesquisa do Estado
do Rio de Janeiro, the SR2-UERJ and the Coordena{\c{c}}{\~{a}}o de
Aperfei{\c{c}}oamento de Pessoal de N{\'\i}vel Superior (CAPES)
for financial support.

\end{document}